\documentclass[accepted=2018-04-23,aps,pre,twocolumn,floatfix,superscriptaddress,showpacs]{quantumarticle}

\usepackage{amsfonts}
\usepackage{amsmath}
\usepackage{amssymb}
\usepackage{mathtools}
\usepackage{graphicx}
\usepackage{color}
\usepackage{bbm}
\usepackage[hidelinks]{hyperref}
\hypersetup{
     colorlinks   = true,
     citecolor    = blue,
     linkcolor    = blue
}
\usepackage{subfigure}
\usepackage{ulem}

\newcommand{\tr}{{\rm Tr}}

\newcommand{\ignore}[1]{}
\newcommand{\nobibentry}[1]{{\let\nocite\ignore\bibentry{#1}}}

\newcommand{\bibfnamefont}[1]{#1}
\newcommand{\bibnamefont}[1]{#1}
\newcommand{\average}[1]{\left<#1\right>}
\newcommand{\ket}[1]{\left\vert#1\right\rangle}
\newcommand{\bra}[1]{\left\langle#1\right\vert}

\begin{document}
\title{Fluctuation-dissipation theorem for non-equilibrium quantum systems}

\author{Mohammad Mehboudi}\affiliation{Departament de F\'isica, Universitat Aut\`onoma de Barcelona - E08193 Bellaterra, Spain}
\affiliation{ICFO-Institut de Ciencies Fotoniques, The Barcelona Institute of
	Science and Technology, 08860 Castelldefels (Barcelona), Spain}
\author{Anna Sanpera}
\affiliation{Departament de F\'isica, Universitat Aut\`onoma de Barcelona - E08193 Bellaterra, Spain}
\affiliation{ICREA, Psg. Llu\' is Companys 23, 08001 Barcelona, Spain}
\author{Juan M.R. Parrondo}
\affiliation{Departamento de F\'isica At\'omica, Molecular y Nuclear and GISC, Universidad Complutense Madrid, 28040 Madrid, Spain}

\begin{abstract}
	The fluctuation-dissipation theorem (FDT) is a central result in statistical physics, both for classical and quantum systems. It establishes a relationship between the linear response of a system under a time-dependent perturbation and time correlations of certain observables in equilibrium. Here we derive a generalization of the theorem which can be applied to any Markov quantum system and makes use of the symmetric logarithmic derivative (SLD). There are several important benefits from our approach. First, such a formulation clarifies the relation between classical and quantum versions of the equilibrium FDT. Second, and more important, it facilitates the extension of the FDT to arbitrary quantum Markovian evolution, as given by quantum maps. Third, it clarifies the connection between the FDT and quantum metrology in systems with a non-equilibrium steady state.
	
\end{abstract}

\pacs{05.30.−d, 05.70.Ln, 03.65.Yz, 03.65.-w, 03.67.−a, 06.20.-f}
\maketitle

The first version of the fluctuation-dissipation theorem (FDT) was derived by Callen and Welton \cite{PhysRev.83.34} and subsequently generalized by Kubo \cite{Kubo57,Kubo66} in the context of linear response theory. Since then, it has been a crucial tool to investigate physical properties, such as transport, energy absorption and susceptibilities, of systems close to thermal equilibrium \cite{Cloizeaux,Jensen,Marconi}. 
More recently, it has been proved useful to assess the multipartite entanglement of complex quantum systems at thermal equilibrium \cite{Hauke2016}, and out of thermal equilibrium \cite{pappalardi2017multipartite}. The usefulness of the FDT in parameter estimation and other related metrology problems is the subject of multiple studies~\cite{PhysRevA.94.062316,PhysRevE.76.022101,doi:10.1142-S0217979210056335,PhysRevA.94.042121}.

Despite the fact that the FDT is so widely used, the standard FDT applies only to small perturbations around thermal equilibrium states \cite{Kubo57,Kubo66,Cloizeaux,Jensen}. There has been an intense activity in the last years to generalize the FDT to classical systems far from equilibrium \cite{Agarwal1972,Seifert,Marconi,1742-5468-2011-10-P10025,Chetrite2011} or to verify it experimentally \cite{PhysRevLett.103.040601}, and more recently to quantum systems \cite{Chetrite2012,PhysRevX.6.041031}. Two main strategies have been followed in this pursuit. The first one looks for correction terms in the original equilibrium FDT \cite{Maess,Sasa}, whereas the second keeps the very mathematical structure of the theorem by redefining the magnitude conjugated to an external parameter \cite{PhysRevLett.103.090601,Seifert1}.

Here, we adopt the second strategy to prove a FDT for generic quantum Markovian systems. The key point in our derivation is the use of the symmetric logarithmic derivative (SLD), $\Lambda_{\lambda}$, of a density matrix $\rho_\lambda$ depending on a real parameter $\lambda$, defined as: 
\begin{eqnarray}\label{eq:sld1}
(\Lambda_{\lambda}\rho_{\lambda}+\rho_{\lambda}\Lambda_{\lambda})\equiv 2\left.\frac{\partial}{\partial \lambda^{\prime}}\right|_{\lambda^{\prime}=\lambda}~\rho_{\lambda^{\prime}}.
\end{eqnarray}
The SLD is an observable with zero average, $\average{\Lambda_{\lambda}}_{\lambda}={\tr}[\Lambda_{\lambda}\rho_{\lambda}]=0$, as can be easily proved by taking the trace of the above equation. It is intimately related to the quantum Fisher information (QFI), 
${\cal F}_\lambda = \tr \left[ \Lambda_{\lambda}^2\rho_{\lambda}\right]$, which
plays a prominent role in metrology, since the uncertainty of any unbiased observable $A$ (i.e. with $\average{A}_{\lambda}=\lambda$), satisfies the
Cr\'amer-Rao bound $\mathrm{Var} (A)_{\lambda}\geq 1/{\cal F}_\lambda$ \cite{PhysRevLett.72.3439,BRAUNSTEIN1996135,doi:10.1142/S0219749909004839,Giovannetti2011,1751-8121-47-42-424006,Giovannetti1330,PhysRevLett.96.010401}. In this paper we show that the SLD provides a novel definition of an observable conjugated to an external parameter, which is extremely useful to derive a completely general FDT for quantum Markov systems and to relate previous versions of the FDT for classical and quantum systems.

We start by applying the SLD to the simplest case of a fluctuation-dissipation relation for the static susceptibility.
Consider a quantum system whose density matrix $\rho_\lambda$ depends on an external parameter $\lambda$. Taking $\rho_0$ as a reference state, we are interested on the change of the expected value of a generic observable $B$ under a small change in $\lambda$. More precisely, for small $\lambda$, the expected value of a generic observable $B$ can be written as:
\begin{equation}\label{eq:staticlr0}
\langle B\rangle_\lambda \equiv {\rm Tr}[B\rho_\lambda ]\simeq \langle B\rangle_0+\chi_B^s\lambda,
\end{equation}
where
\begin{equation}
\chi_{B}^s \equiv \left.{\partial_\lambda}\right|_{\lambda=0} \langle B\rangle_\lambda={\rm Tr}\left[B \left.{\partial_\lambda}\right|_{\lambda=0} \rho_\lambda\right],
\end{equation}
is the \textit{static susceptibility} of observable $B$.
Using the SLD, the derivation of a fluctuation-dissipation relation is straightforward:
\begin{align}
\chi_{B}^s =\frac{1}{2}{\rm Tr}\left[ B(\Lambda_0\rho_0+\rho_0\Lambda_0)\right]=\frac{1}{2}\langle B\Lambda_0 +\Lambda_0 B\rangle_0 \label{eq:staticfdr}
\end{align}
which is the symmetrized correlation between observables $B$ and $\Lambda_0$, since $\langle \Lambda_0\rangle_0=0$. 
In the Appendix~\ref{app A} we show that Eq.~\eqref{eq:staticfdr}, when particularized to a thermal state $\rho_\lambda=e^{-\beta (H_0-\lambda A)}/Z(\lambda)$, with $\beta = 1/(kT)$ and $Z(\lambda)\equiv {\rm Tr}\left[e^{-\beta(H_0-\lambda A)}\right]$ 
and expressed in the eigenbasis of $H_0$, yields the standard fluctuation-dissipation relation.

We now turn to the case of a generic Markov evolution given by the composition of completely positive and trace preserving (CPTP) maps. Let $\xi_\lambda(\rho)$ be a CPTP map that depends on a parameter $\lambda$. We assume that each map $\xi_\lambda$ has an invariant state $\pi_\lambda$, i.e., $\xi_\lambda(\pi_\lambda)=\pi_\lambda$. We study a small time-dependent perturbation $\lambda(t)$ affecting the invariant state $\pi_0$. More precisely, we consider the evolution, in discrete time steps $t=1,2,\dots$,
\begin{equation}\label{eq:evol1}
\rho(t)=\xi_{\lambda(t)}\circ\xi_{\lambda(t-1)}\circ\dots\circ\xi_{\lambda(1)}(\pi_0).
\end{equation}
The linear response of an observable $B$ can be written as:
\begin{equation}\label{eq:lr0}
\langle B(t)\rangle =\langle B\rangle_0+\sum_{t'=1}^{t} \phi_B(t-t')\lambda(t')
\end{equation}
where $\langle B\rangle_0={\rm Tr}[\pi_0B]$, $\langle B(t)\rangle\equiv {\rm Tr}[\rho(t)B]$, and $\phi_B(t-t')$ is the {\it response function} of the observable $B$ under the perturbation $\lambda$.

One can extend the above definition to the case of maps acting for a short time $\Delta t$. In the continuous limit, $\Delta t\to 0$, the sum in \eqref{eq:lr0} is replaced by the integral
\begin{equation}\label{eq:lr0cont}
\langle B(t)\rangle =\langle B\rangle_0+\int_{0}^t dt'\, \phi_B(t-t')\lambda(t').
\end{equation}
The {\it generalized susceptibility} is defined as the Fourier transform of the response function ($\phi_B(t)$ is assumed to vanish for $t<0$ due to causality):
\begin{equation}
\chi_{B}(\omega)=\int_0^\infty dt\, \phi_{B}(t)e^{i\omega t}.
\end{equation}
The generalized susceptibility has interesting properties such as the {\it Kramers-Kronig} relation between the real and imaginary parts $\chi_B(\omega)=\chi_B'(\omega)+i\chi_B''(\omega)$. When the evolution is unitary under the Hamiltonian $H_0-\lambda(t)A$, $\chi_B''(\omega)$ is called absorptive part of the susceptibility \cite{Kubo57,Kubo66,Cloizeaux,Jensen}, since the energy absorbed by the system due to the perturbation is proportional to $\chi_B''(\omega)$.

The static susceptibility $\chi_B^s$ can be related to the response function by considering a constant perturbation $\lambda(t)=\lambda$ for $t\geq 0$ \cite{Kubo57,Kubo66,Cloizeaux,Jensen}. In this case $\langle B(t)\rangle \to \langle B\rangle_\lambda$ when $t\to\infty$ and Eq.~\eqref{eq:lr0cont} 
implies that the static susceptibility is the integral of the response function or, equivalently, the generalized susceptibility at zero frequency:
$\chi_{B}^s=\chi_B(\omega=0)$.

To obtain the FDT we calculate $\rho(t)$ up to linear terms in $\lambda$. For that, we write $\xi_\lambda =\xi_0+\lambda\xi_1+\dots$ (notice that $\xi_1$ is not a CPTP map) and the invariant state as $\pi_\lambda = \pi_0+\lambda\pi_1+\dots$ The invariance of $\pi_\lambda$ under the map $\xi_\lambda$ implies:
\begin{equation}\label{eq:pil}
\xi_1(\pi_0)+\xi_0(\pi_1)=\pi_1
\end{equation}
and the SLD of $\pi_\lambda$ with respect to $\lambda$ at $\lambda=0$ obeys
\begin{equation}\label{eq:sldpi}
2\pi_1=\Lambda_0\pi_0+\pi_0\Lambda_0.
\end{equation}

Expanding the evolution equation \eqref{eq:evol1} up to linear terms, we obtain
\begin{eqnarray}
\rho(t) &=&\xi_0^t(\pi_0)+\sum_{t'=1}^{t} \lambda(t')\,\xi_0^{t-t'}\circ \xi_1\circ \xi_0^{t'-1}(\pi_0) \nonumber \\
&=&\pi_0 +\sum_{t'=1}^{t}  \lambda(t')\,\xi_0^{t-t'}\circ \xi_1(\pi_0)
\end{eqnarray}
where we have used the invariance of $\pi_0$ under $\xi_0$. The expected value of $B$ is
\begin{equation}\label{eq:bt}
\langle B(t)\rangle =\langle B\rangle_0+\sum_{t'=0}^{t} {\rm Tr}\left[ B\, \xi_0^{t-t'}\circ \xi_1(\pi_0)\right]\lambda(t').
\end{equation}
Comparing \eqref{eq:bt} with \eqref{eq:lr0}, we immediately get
\begin{equation}
\phi_B(t)={\rm Tr}\left[B\,\xi_0^{t}\circ \xi_1(\pi_0)\right].
\end{equation}
Using \eqref{eq:pil} and \eqref{eq:sldpi}:
\begin{align}
\phi_B(t)&= {\rm Tr}\left[B\,\xi_0^{t}\circ(\pi_1-\xi_0(\pi_1))\right] \nonumber\\
&= {\rm Tr}\left[B\xi_0^{t}(\pi_1)-B\xi_0^{t+1}(\pi_1)\right] 
= -{\rm Tr}\left[\Delta B(t)\pi_1\right] 
\nonumber \\
&=-\frac{1}{2}{\rm Tr}\left[\Delta B(t)(\Lambda_0\pi_0+\pi_0\Lambda_0)\right]\nonumber \\
&=-\frac{1}{2}\langle \Delta B(t)\Lambda_0 + \Lambda_0 \Delta B(t)\rangle_0\label{eq:fdr}
\end{align}
where $\Delta B(t)=B(t+1)-B(t)$ and $B(t)=\tilde\xi_0^{t}(B)$ is the evolution of the observable $B$ in the generalized Heisenberg picture for quantum maps \cite{breuer2002theory,Nielsen:2011:QCQ:1972505,wilde2013quantum}. Here $\tilde\xi_0(\cdot)$ is the adjoint map (not necessarily trace preserving) with respect to the scalar product between operators given by the trace, i.e., ${\rm Tr}[A\xi_0(B)]={\rm Tr}[\tilde\xi_0(A)B]$, for all pair of operators $A$ and $B$.

The fluctuation-dissipation relation for the static case \eqref{eq:staticfdr} is recovered from \eqref{eq:fdr} if the correlations between $B(t)$ and $\Lambda$ vanish in the limit $t\to\infty$:
\begin{eqnarray}
\chi_{B}^s &=&\sum_{t'=0}^{\infty}\phi_B(t')\nonumber \\
&=&-\frac{1}{2}\lim_{t\to\infty}\left(\langle B(t)\Lambda_0+\Lambda_0 B(t)\rangle_0+\langle B\Lambda_0+\Lambda_0 B\rangle_0\right)\nonumber\\
&=& \frac{1}{2}\langle B\Lambda_0+\Lambda_0 B\rangle_0.
\end{eqnarray}
Finally, the continuous-time version of theorem \eqref{eq:fdr} is
\begin{equation}
\phi_B(t)=-\frac{1}{2}\,d_t\,\langle  B(t)\Lambda_0 + \Lambda_0 B(t)\rangle_0.\label{eq:contfdr}
\end{equation}
Eqs.~\eqref{eq:fdr} and \eqref{eq:contfdr} are our main result.
These results are the quantum counterparts of the non-equilibrium classical FDT derived by Agarwal in \cite{Agarwal1972} and revived recently in \cite{PhysRevLett.103.090601,Seifert1}. Notice that, in the classical scenario, the conjugate variable is defined as the derivative of the logarithm of the steady state probability distribution. On the quantum scenario, the non-commutativity of observables does not allow to simply replace the classical conjugate variable with the derivative of the logarithm of the density matrix. Nevertheless, the choice of the SLD observable solves this non-commutativity issue.
In the Appendix~\ref{app B} we prove that the latter expression \eqref{eq:contfdr}, when particularized to Hamiltonian dynamics and equilibrium states of the form $\pi_\lambda=e^{-\beta(H_0-\lambda A)}/Z(\lambda)$, is completely equivalent to the standard Kubo formula \cite{Kubo57,Kubo66,Cloizeaux,Jensen}:

\begin{equation}
\phi_{BA}(t)=\frac{i}{\hbar}{\rm Tr}\big[[A,\pi_0]B(t)\big]=\frac{i}{\hbar}\langle [B(t),A]\rangle.
\end{equation}
It is worth it to point out that, in the quantum case, the Kubo formula does not yield a simple FDT; more precisely, the response function cannot be expressed in terms of the time derivative of a two-time correlation in equilibrium. Such a relationship can only be derived in the frequency domain for the absorptive part of the generalized susceptibility and the Fourier transform of the correlation (see \cite{Kubo66,Cloizeaux,Jensen} and the Appendix~\ref{app B}). On the contrary, our generalized FDT, namely Eqs.~\eqref{eq:fdr} and \eqref{eq:contfdr}, expresses the response function in terms of correlations in the time domain and can be equally applied to classical and quantum systems. This uniform formulation is possible due to the introduction of the SLD. In the classical case, the SLD coincides with the normal derivative and consequently, for a thermal state with Hamiltonian $H_0-\lambda A$, the SLD is $-\beta (A-\langle A\rangle)$, whereas for a quantum system with $[H_0,A]\neq 0$, the SLD yields a nontrivial conjugated variable as shown below. We highlight the usefulness of our FDT for quantum metrology explicitely through the examples that follow. However, let us remark that such a link holds for any map that fits our framework. 

We illustrate the new generalized FDT with a simple example consisting of
two harmonic oscillators with a modulated interaction. The Hamiltonian reads:
\begin{eqnarray}
\label{2ocillators}
H&=&H_0+H_J\cr 
&=&\omega   a_1^{\dagger}a_1+(\omega+\delta)a_2^{\dagger}a_2-J(t)(a_2^{\dagger}a_1+a_1^{\dagger}a_2),\nonumber\\
\end{eqnarray}
where $a_i$ and $a_i^\dagger$ are the ladder operators of $i$-th oscillator and $\delta$ denotes the frequency detuning between the oscillators. We assume $|J(t)|\ll\omega$ so that linear response theory holds at any time. 
First, we consider the two harmonic oscillators placed in a bath at temperature $T$, and examine the response of the system to a perturbation $J(t)=J_0$ around the thermal equilibrium state $\rho_0=\mathrm{exp}(-\beta H_{0})/Z$.
Classically, the susceptibility is defined as $\partial_J\average{A}_J$, with $A=\partial_{J}H=-(a_2^{\dagger}a_1+a_1^{\dagger}a_2)$ being 
the conjugate variable. Notice that if the detuning is zero, then
$[H_0,A]=0$, and the SLD reads $\Lambda_J(\delta=0)=-\beta (A-\average{A}_J)$ \cite{1367-2630-19-10-103003}. On the contrary, any $\delta\neq 0$ forces that $[H_0,A]\neq 0$, and the SLD cannot be anymore identified trivially as the conjugate variable $A$ \cite{PhysRevA.94.042121}. 
For a finite detuning, the SLD takes the form $\Lambda_J(\delta)=C(\delta)\Lambda_J(0)$ (see \cite{monras2013phase,PhysRevA.89.032128}).
This additional coefficient arises from the non-commutativity between $H_0$ and $A$, and reads:
\begin{align}
C(\delta)\equiv\frac{\tanh(\delta/2T)}{\delta/2T}.
\end{align}  
The use of the SLD allows us to show that the QFI is:
\begin{align}
\mathcal{F}_J(\delta)
=C^2(\delta)\,\beta^2\,\mathrm{Var}(A)_J.
\end{align}	
The additional coefficient is bounded, $0< C(\delta)\leq 1$. It has a maximum at $\delta=0$, and then it decreases monotonically with the detuning, therefore, the precision on the estimation of $J$ decreases as the detune increases.
  
We now address the problem of a time-dependent modulation of $J(t)$ in a non-equilibrium environment induced by two thermal baths at different temperatures $T_1$ and $T_2$. 
A master equation that has been widely used to describe the reduced state of the oscillators consists of a Linblad equation \cite{Campbell2016,breuer2002theory,wiseman2009quantum,gardiner2004quantum}
\begin{eqnarray}
\dot{\rho}(t)=-i[H,\rho(t)]+\sum_{i=\{1,2\}}\mathcal{D}_i[\rho(t)],
\label{Master-Interaction-Double}
\end{eqnarray}
with two independent dissipators $\mathcal{D}_{i}$ 
\begin{eqnarray}
\mathcal{D}_i[\rho(t)]&=\gamma(N_i+1)\left(a_i\rho(t) a_i^{\dagger}-1/2\{a_i^{\dagger}a_i,\rho(t)\}\right)\nonumber\\
&+\gamma N_i\left(a_i^{\dagger}\rho(t) a_i-1/2\{a_ia_i^{\dagger},\rho(t)\}\right).
\label{Double_Dissipater}
\end{eqnarray} 

Here $N_i=(\exp[\beta_i\omega_i]-1)^{-1}$ is the mean occupation number of the $i$-th oscillator, and $\gamma$ is the dissipation rate, which is assumed to be equal for both oscillators. The equation is only valid for small $J(t)$---compared to $\omega$, i.e., the system's energy scale---and $J(t)<\delta$ \cite{gonzalez2017testing}. In particular, it does not predict the thermalization of the full two-oscillator system for finite $J$ when $T_1=T_2$. Obviously, for $J(t)=0$, both oscillators evolve independently reaching a stationary state in which each oscillator is at thermal equilibrium with its own bath:
\begin{align}
\rho_0=\frac{\mathrm{e}^{-\beta_1 H_1}}{\mathrm{Tr}[\mathrm{e}^{-\beta_1 H_1}]} \otimes
\frac{\mathrm{e}^{-\beta_2 H_2}}{\mathrm{Tr}[\mathrm{e}^{-\beta_2 H_2}]},
\label{initial-double-rho}
\end{align}
where $H_i$ denotes the free Hamiltonian of the $i$-th oscillator. 
Since the model under study is quadratic in creation and annihilation operators, one may equivalently describe it by means of its covariance matrix (CM) which contains only first and second moments (i.e., Gaussian), notably simplifying the calculations. The latter is described with the help of the quadratures
\begin{align}
x_i =\frac{1}{\sqrt{2}}(a_i^{\dagger}+a_i),\quad
p_i =\frac{i}{\sqrt{2}}(a_i^{\dagger}-a_i)
\label{eq-quadratures-definition}
\end{align}
which satisfy the standard commutation relation $[x_j,p_k]=i\delta_{jk}$.
In turn, the matrix elements of the CM are defined as follows: 
$\sigma_{jk}\equiv\average{R_jR_k+R_kR_j}/2-\average{R_j}\average{R_k}$, with $R_j\in\{x_1,x_2,p_1,p_2\}$.
The CM corresponding to the stationary state of the master equation \eqref{Master-Interaction-Double} is provided in the Appendix~\ref{app C}, where we also find (i) the SLD, and (ii) the time evolution of all the quadratic observables.
Note that according to Eq.~\eqref{eq:contfdr}, (i) and (ii) are the two key elements required for evaluation of $\phi_B(t)$. Specifically, the SLD writes as:
\begin{align}
\mathrm{\Lambda}_0&=c_1(x_1x_2+p_1p_2)+c_2(x_1p_2-p_1x_2),
\end{align}
with $c_1$ and $c_2$ being real numbers.
Therefore, the SLD is a non-local operator, hence the response of any local observable vanishes.

To proceed further, let us consider $J(t)=J_0(1-\cos\nu\,t)$. 
\begin{figure*}[]
	\includegraphics[width=0.44\linewidth]{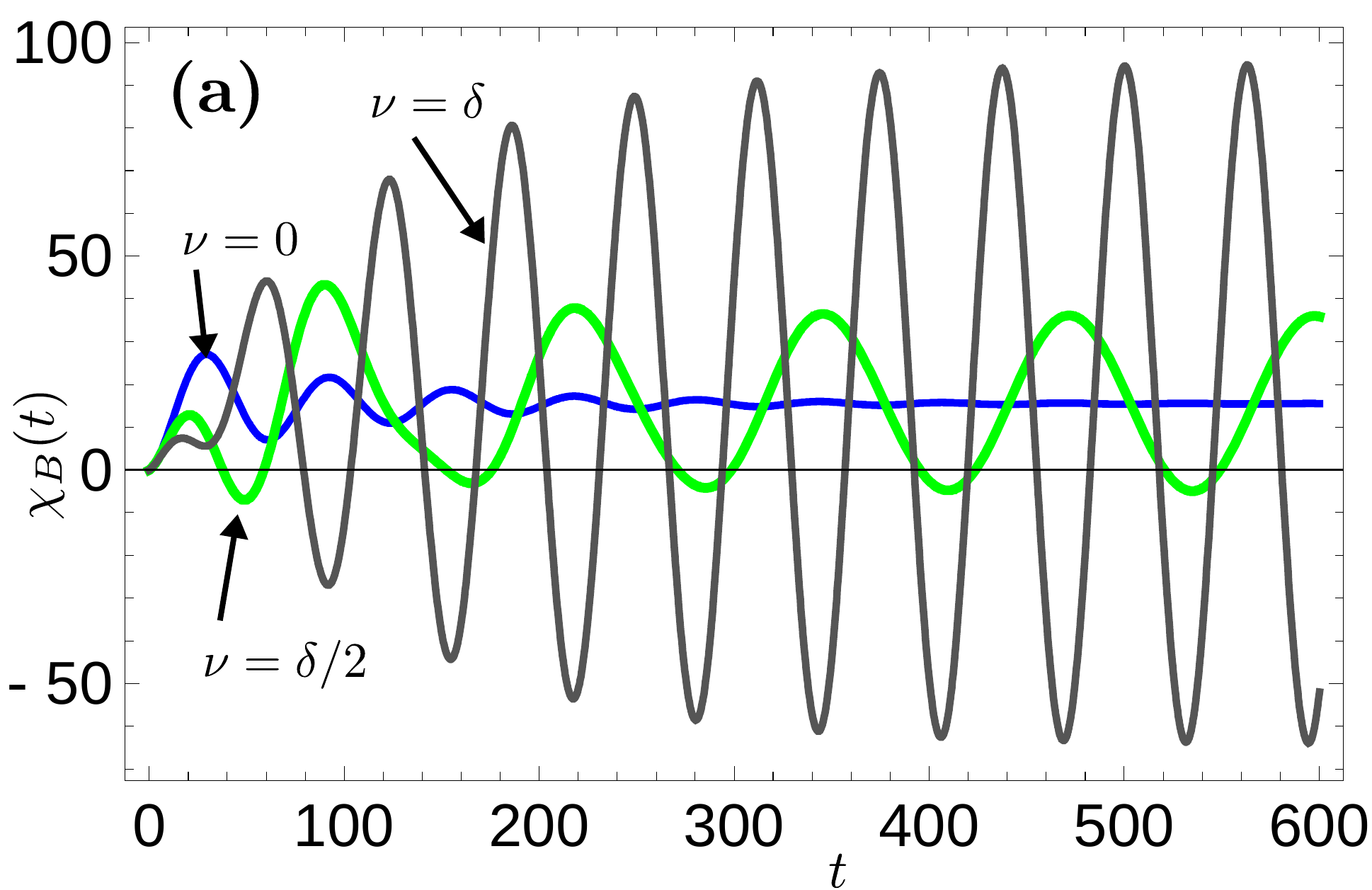}
	\hspace{1cm}
	\includegraphics[width=0.44\linewidth]{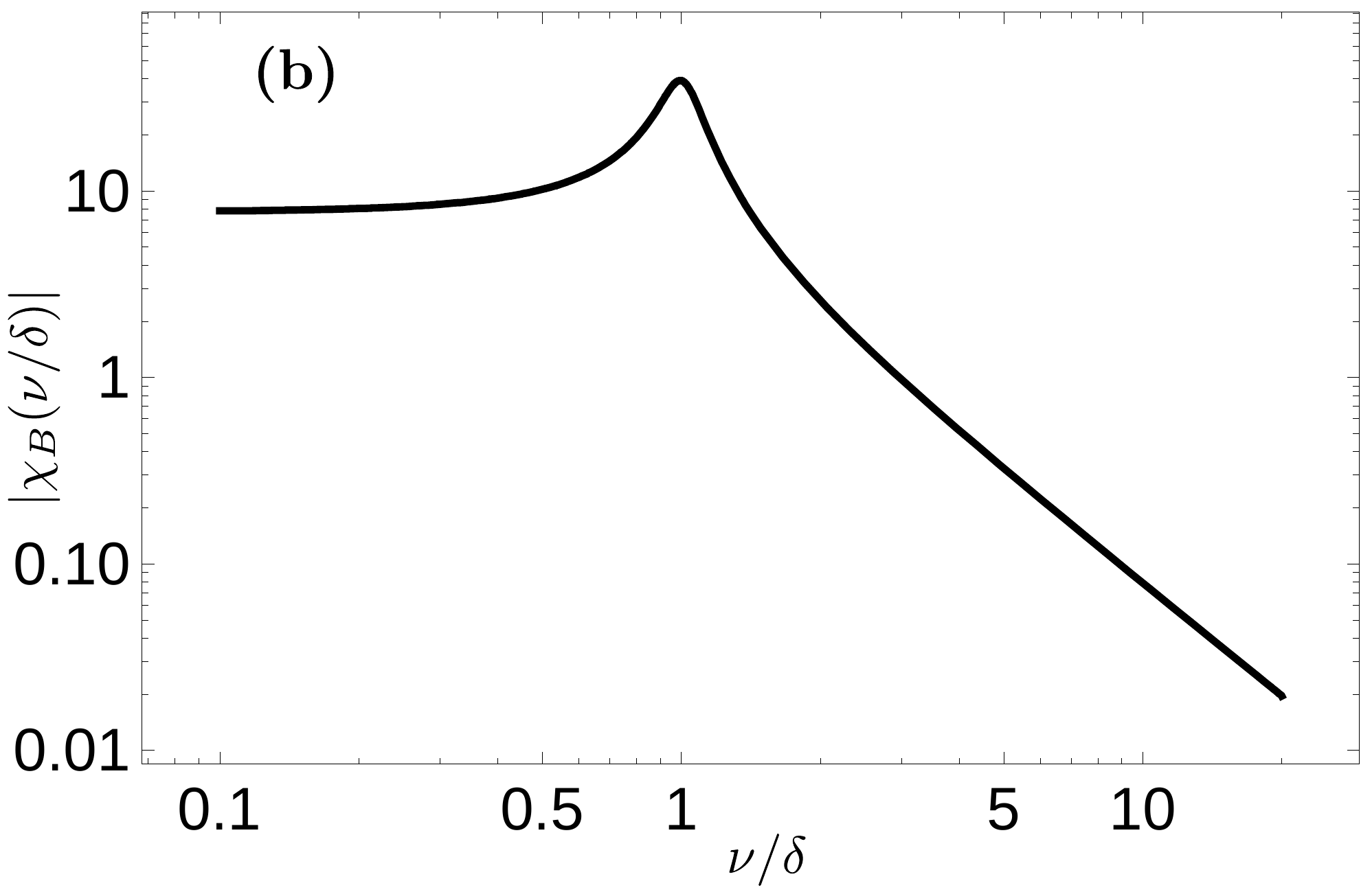}
	\caption{(color online) 
		{(a)} Time dependent susceptibility of the observable $B=x_1x_2$ versus time, and for three different modulation frequencies. {(b)} The dynamical susceptibility as a function of the modulation frequency. See the text for details.
		In both panels we set $\omega=1$, $\delta=0.1$, $T_1=1$, $T_2=2$, and $\gamma=0.01$.}
	\label{Fig-Double-Well_Linear_Response_resonance}
\end{figure*}
It is convenient to define the \textit{time dependent} counter part of the static susceptibility as follows:
\begin{align}
\chi_B(t)\equiv\partial_{J_0}\left.\right|_{J_0=0}\average{B(t)},
\end{align}
which quantifies the deviations of the observable from its initial value, and does not depend on $J_0$.
Note that for a constant perturbation (i.e., $\nu=0$) we recover $\chi_B(t\to\infty)=\chi_B^s$. 
In Fig.~\ref{Fig-Double-Well_Linear_Response_resonance}~(a)
we depict $\chi_B(t)$ versus time for $B=x_1x_2$ evaluated at three different modulation frequencies $\nu=0$, $\nu=\delta/2$, and $\nu=\delta$.
For the constant perturbation, we observe that at short times $\chi_B(t)$ is oscillating but, as time passes, the system relaxes into a new steady state such that $\average{B(t\to\infty)}=\average{B}_J$.
On the contrary, for $\nu\neq 0$, the system never relaxes to a stationary state. In this case, the time dependent susceptibility can be approximated at sufficiently large times by:
\begin{align}
\chi_B(t)&\approx|~{\chi}_B(\nu)~|\cos(\nu t+\alpha),
\label{eq:approximation}
\end{align}
where $\alpha=\arctan\big(\mathrm{Re}~{\chi}_B(\nu)/\mathrm{Im}~{\chi}_B(\nu)\big)$.
Effectively, the generalized suceptibility $|{\chi}_B(\nu)|$, maps onto the amplitude of the oscillations of the time dependent susceptibility, and its dependence on $\nu$ is illustrated in Fig.~\ref{Fig-Double-Well_Linear_Response_resonance}~{(b)}. One can see that the dynamical susceptibility is a flat function for $\nu\ll\delta$, increases sharply as $\nu$ turns out to be resonant with the detuning, and strongly decreases afterwards. 
A similar behaviour is observed for any other nonlocal observable $B=\{x_1p_2;p_1x_2;p_1p_2\}$.
\begin{figure}
	\includegraphics[width=0.95\linewidth]{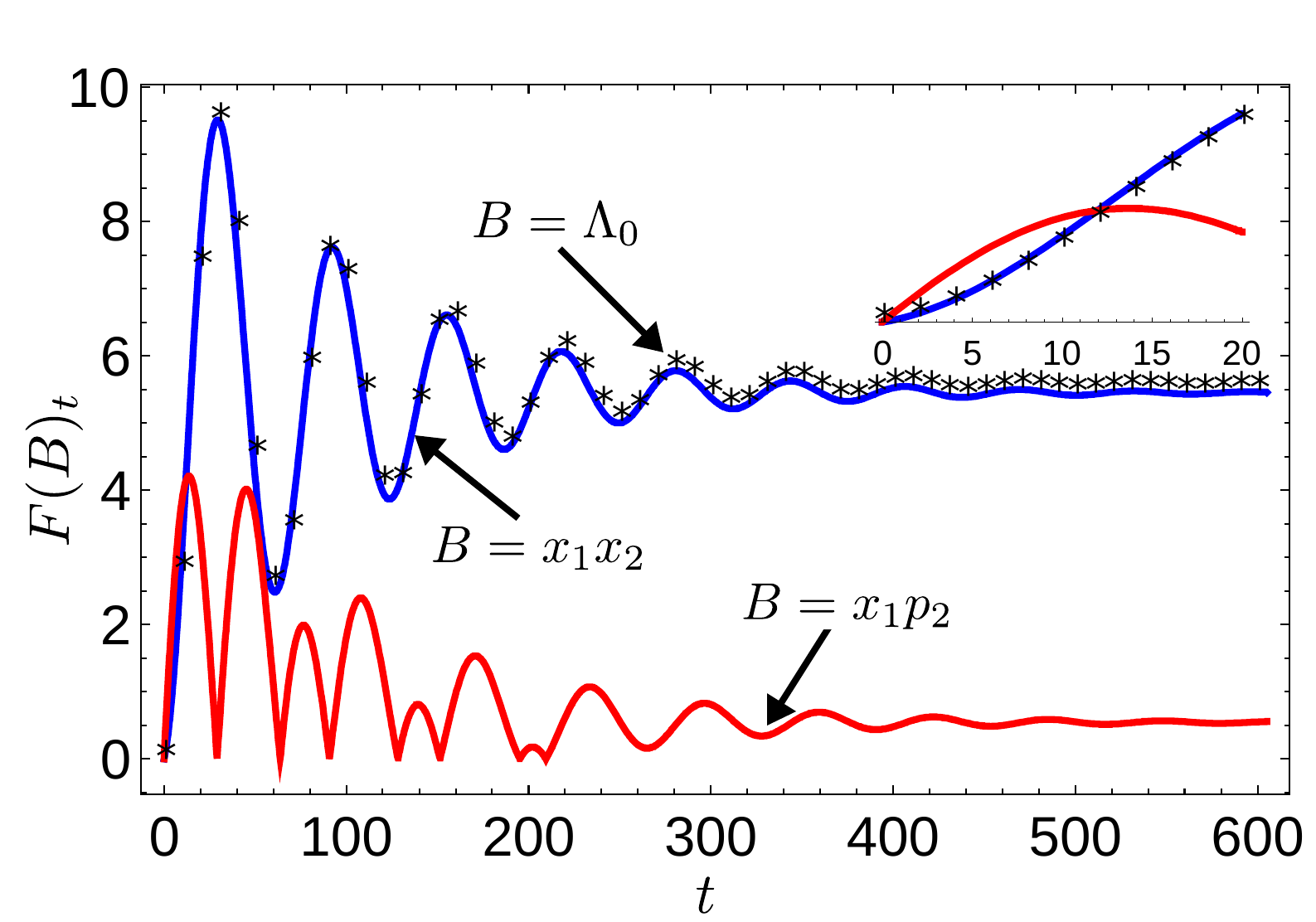}
	\caption{(color online) The sensitivity, $F(B)_t$ as a function of time for $B =\{x_1x_2; x_1p_2; \Lambda_0\}$ for a constant perturbation $J(t)=J_0$ and otherwise same parameters as in Fig.~\ref{Fig-Double-Well_Linear_Response_resonance}.}
	\label{Fig-Double-Well-Constant}
\end{figure}

Futhermore, the metrological character of the FDT when expressed through the SLD, can be seen by identifying which is the best measurement to detect the smallest interaction coupling $J(t)=J_0$. 
If we have no prior knowledge about $J_0$, 
linear response theory ensures that $Tr[B(\rho(t)-\rho_0)]=J_0\,\chi_B(t)$. Unavoidable, an expectation value bears a statistical error, $\sqrt{\mathrm{Var}(B)_0/n}$, by repeating the measurement $n$ times\footnote{In fact the statistical error at measurement time is $\mathrm{Var}(B)_t$, but since we only keep the leading order, in the linear response regime we safely replace it by $\mathrm{Var}(B)_0$.}. Thus, in order to infer the value of the perturbation, the linear response of the system must be larger than the statistical error, i.e., $|J_0 \chi_B(t)|>\sqrt{\mathrm{Var}(B)_0/n}$, setting a lower bound on the smallest perturbation that one may detect by measuring $B$. The inverse of this lower bound
defines the sensitivity $F(B)_t$ of the observable $B$ to the perturbation $J_0$:
\begin{align}
F(B)_t\equiv|\chi_B(t)|/\sqrt{\mathrm{Var}(B)_0/n}.
\label{Sensitivity}
\end{align}
Notice that the above definition is not restricted to the model discussed here. It can be applied for estimation of a generic parameter which is parametrized through a divisible map as in \eqref{eq:evol1}.
For a constant perturbation, as $t\rightarrow\infty$,
$\chi_B(t\to\infty)\approx\mathrm{Corr}(B,\mathrm{\Lambda}_{0})_0$. Substituting this in Eq.~\eqref{Sensitivity}, and using the Cauchy-Schwartz inequality, the upper bound on the sensitivity of $B$ at $t\rightarrow\infty$ is given by: 
\begin{align}
F(B)_{\infty}\leq\sqrt{n\mathrm{Var}(\Lambda_{0})_0}=\sqrt{n\mathcal{F}_{0}},
\label{Optimal-Sensitivity}
\end{align}
with $\mathcal{F}_{0}$
being the QFI evaluated at $J_0=0$. 
The QFI has been previously used to quantify the ultimate precision of parameter estimation in non-equilibrium steady states of spin models \cite{PhysRevA.90.062130}.
The bound \eqref{Optimal-Sensitivity} is saturated by performing a measurement in the $\Lambda_{0}$ basis. In Fig.~\ref{Fig-Double-Well-Constant}, 
we depict $F(B)_t$ for three observables $B=\{\Lambda_0;x_1x_2;x_1p_2\}$. Although no observable can overperform the sensitivity of the SLD at $t\to\infty$, at shorter times this fails to be the case as shown in the inset of Fig.~\ref{Fig-Double-Well-Constant}. More interestingly, the maximum value of $F(B)_t$ is not necessarily achieved at $t\rightarrow \infty$. A similar behavior is observed for a time-dependent perturbation $J(t)=J_0(1-\cos \nu t)$. Although in this case
inequality~\eqref{Optimal-Sensitivity} does not apply, the FDT as expressed in Eq.~\eqref{eq:contfdr} links $\Lambda_0$ with the response function. From the oscillatory character of the time dependent susceptibility, the optimal times to best estimate $J_0$ are given by Eq.~\eqref{eq:approximation}.

In summary, we have presented a novel formulation of the Fluctuation-Dissipation Theorem (FDT) in terms of the symmetric logarithmic derivative (SLD), which is completely general to describe the effect of a perturbation on a quantum systems, within the linear response. Such a formulation presents some clear advantages. First, it unifies FDT for classical and quantum systems. Second, it can be straightforwardly applied to any Markovian evolution by means of quantum maps. This permits the extension of the FDT to non-equilibrium dynamics. Third, it provides an explicit connection between the susceptibility of an observable to an external perturbation and the figure of merit in quantum metrology. 

Our FDT can be used to generalize the detection of multipartite entanglement to the non-equilibrium steady states (NESS). Although at thermal equilibrium, or after quenching a thermal state, such relations are known~\cite{Hauke2016,pappalardi2017multipartite}, there are less studies for a general NESS state. In this regard, the sensitivity measure Eq.~\eqref{Sensitivity}---which is a lower bound on the QFI~\cite{1751-8121-47-42-424006,PhysRevLett.120.020506}---can be used to detect multipartite entanglement.
Finally, our results prompt an interesting open question on the relationship between the quantum versions of FDT and fluctuation theorems (FT). Needless to mention that, the quantum versions of FT have been subject of extensive studies, that have produced a rich literature on the subject~\cite{RevModPhys.83.771,RevModPhys.81.1665,PhysRevX.8.011019}. The relationship between FDT and FT has been made clear for classical systems. On the other hand, the FT for quantum CPTP maps requires a condition that is not necessary in our derivation of the FDT, namely, the Kraus operators of the map must be ladder operators in a relevant basis: the eigenbasis of the instantaneous stationary state $\pi_0$ for generic CPTP maps \cite{PhysRevE.92.032129} or, for periodically driven systems in contact with equilibrium reservoirs, Floquet eigenstates \cite{1367-2630-17-5-055002} or displaced energy eigenstates \cite{PhysRevE.85.031110}. Our FDT suggests that one could obtain a more general FT by using the SLD.\\
\begin{acknowledgments}
	We acknowledge financial support from the Spanish MINECO projects FIS2013-40627-P, FIS2013-46768, FIS2014-52486-R (AEI/FEDER EU), QIBEQI FIS2016-80773-P, Severo Ochoa SEV-2015-0522, and the Generalitat de Catalunya CIRIT (2014-SGR-966, 2014-SGR-874), and the Generalitat de Catalunya (CERCA Programme) and Fundaci\'{o} Privada Cellex.
	M.M. acknowledges financial support from E.U. under the project TherMiQ.
\end{acknowledgments}
%
\onecolumngrid
\appendix
\section{Quantum systems in thermal equilibrium}
\label{app A}

A relevant particularization of the static fluctuation dissipation relation Eq.~\eqref{eq:staticfdr} is the case of a quantum system with Hamiltonian $H_0-\lambda A$ at thermal equilibrium. In such a case the density matrix is $\rho_\lambda=e^{-\beta (H_0-\lambda A)}/Z(\lambda)$, where $\beta = 1/(k_BT)$ is the inverse temperature and $Z(\lambda)\equiv {\rm Tr}\left[e^{-\beta(H_0-\lambda A)}\right]$ is the partition function of the system.
The equilibrium static susceptibility of an arbitrary observable $B$ under the perturbation $\lambda A$ is denoted as $\chi_{B}^s$ and obeys Eq.~\eqref{eq:staticfdr}. 
Furthermore, when the SLD is expressed in 
the eigenbasis of the unperturbed Hamiltonian $H_0\ket{n}=E_n\ket{n}$, one can obtain some interesting relationships for the equilibrium static susceptibility. It is convenient to rewrite the SLD using the Feynman's formula:
\begin{equation}\label{eq:feynman}
\left.\frac{\partial}{\partial \lambda}\right|_{\lambda=0} e^{-\beta (H_0-\lambda A)}=\beta\int_0^1 ds\,e^{-\beta H_0 (1-s)}A e^{-\beta H_0 s}.
\end{equation}
\\
In the eigenbasis of $H_0$, the formula reads:
\begin{align}
\bra{n}\left.\frac{\partial}{\partial \lambda}\right|_{\lambda=0} e^{-\beta ( H_0-\lambda A)}\ket{m}= \beta \bra{n}A\ket{m} \int_0^1 ds\,e^{-\beta [E_n (1-s)+E_ms]}
=\bra{n}A\ket{m} \frac{e^{-\beta E_m}-e^{-\beta E_n}}{E_n-E_m},
\end{align}
if $E_n\neq E_m$. Otherwise, the matrix element is $\beta\bra{n}A\ket{m}e^{-\beta E_n}$. Therefore:
\begin{align}
\bra{n}\left.\frac{\partial}{\partial \lambda}\right|_{\lambda=0} \rho_\lambda\ket{m}  
&=\bra{n}A\ket{m} \frac{p_m-p_n}{E_n-E_m}, \quad\mbox{for $E_n\neq E_m$},\nonumber\\
\bra{n}\left.\frac{\partial}{\partial \lambda}\right|_{\lambda=0} \rho_\lambda\ket{m} &=p_n\left[\beta\bra{n}A\ket{m}-\delta_{mn} \frac{Z'(0)}{Z(0)}\right], \quad\mbox{for $E_n=E_m$}.
\end{align}
Here $p_n=e^{-\beta E_n}/Z(0)$ is the population of level $E_n$ at equilibrium. Using Eq.~\eqref{eq:feynman}:
\begin{eqnarray}
\frac{Z'(0)}{Z(0)}&=&\frac{\beta}{Z(0)}\int_0^1 ds\,{\rm Tr}\left[ e^{-\beta H_0 (1-s)}A e^{-\beta H_0 s}\right]\nonumber
\\
&=& \frac{\beta}{Z(0)} {\rm Tr}\left[e^{-\beta H_0} A\right]\nonumber\\
&=&\beta \langle A\rangle_0.
\end{eqnarray}
Eq.~\eqref{eq:sld1}  can be written as:
\begin{equation}
2\bra{n}A\ket{m} \frac{p_m-p_n}{E_n-E_m}
=(p_n+p_m) \bra{n}\Lambda_0\ket{m},
\end{equation}
for $E_n\neq E_m$. For $E_n = E_m$ by choosing the eigenbasis $\ket{n}$ such that $A$ is diagonal in the eigen-subspaces of $H_0$ we have\footnote{This means that the eigenstates of $H_0$ are chosen such that for any two states $\ket{n},\ket{m}$ with the same energy, the criteria $\bra{n}A\ket{m}\neq 0$ holds only if $m=n$. Therefore, $\bra{n}A\ket{m}=\bra{n}A\ket{m}\delta_{m,n}$.}:
\begin{equation}
2 \beta\left[ \bra{n}A\ket{m}- \langle A\rangle_0\right]p_n\delta_{mn}=2\,p_n \bra{n}\Lambda_0\ket{m},
\end{equation}
Therefore:
\begin{eqnarray}
\bra{n}\Lambda_0\ket{m} &=& 2\frac{\bra{n}A\ket{m}}{p_n+p_m} \frac{p_m-p_n}{E_n-E_m} \quad\mbox{for $E_n\neq E_m$}, \nonumber \\
\bra{n}\Lambda_0\ket{m} &=& \beta \left[ \bra{n}A\ket{m}-\langle A\rangle_0\right]\delta_{mn}  \quad \mbox{for $E_n= E_m$}.\nonumber 
\end{eqnarray}
Using this expression, the susceptibility of $A$ reduces to:
\begin{eqnarray}
\chi_{A}^s &=& \frac{1}{2}\sum_n p_n\bra{n}\Lambda_0~A+A~\Lambda_0\ket{n} \nonumber \\
&=& \frac{1}{2}\sum_{n,m} p_n \big( \bra{n}\Lambda_0\ket{m}\bra{m}A\ket{n}~+~\bra{n}A\ket{m}\bra{m}\Lambda_0\ket{n} \big)\nonumber\\
&=& \frac{1}{2}\sum_{n,m} (p_n+p_m) \bra{n}\Lambda_0\ket{m}\bra{m}A\ket{n} \nonumber\\
&=& \beta \left[ \sum_n p_n |\bra{n}A\ket{n}|^2 -\langle A\rangle_0^2\right] 
+ \sum_{n,m}\frac{p_m-p_n}{E_n-E_m} |\bra{n}A\ket{m}|^2,
\end{eqnarray}
where the last sum runs over all $n$ and $m$ with $E_n\neq E_m$.
One can distinguish the Curie and van Vleck terms in the FDT \cite{Cloizeaux,Jensen}.

It is interesting to consider the observable $\tilde A = \Lambda_0$,
i.e., the SLD of $\rho_\lambda$ at $\lambda=0$. Since $[\tilde A,\Lambda_0]=0$ and $\langle \Lambda_0\rangle_0=0$, the susceptibility of $\tilde A$ is equivalent to the QFI, i.e., we have $\chi_{\tilde A}^s=\mathrm{Var}(\Lambda_0)_0=\mathcal{F}_0$. Therefore, $\tilde A$ identifies the most sensitive observable to the perturbation (rather than $A$), and its sensitivity saturates the Cram\'er-Rao bound, that is given by:
\begin{eqnarray}
\mathcal{F}_0=\chi_{\tilde A}^s &=& \sum_n p_n\bra{n}\Lambda_0^2\ket{n} \nonumber \\
&=& \beta \left[ \sum_n p_n |\bra{n}A\ket{n}|^2 -\langle A\rangle_0^2\right] 
+2\sum_{n,m}\left(\frac{p_m-p_n}{E_n-E_m}\right)^2\,\frac{|\bra{n}A\ket{m}|^2}{p_n+p_m} .
\end{eqnarray}

\section{Kubo relations}
\label{app B}

Our main results Eqs.~\eqref{eq:fdr} and \eqref{eq:contfdr} are FDTs for generic quantum Markov systems. One can recover the familiar Kubo quantum FDT for states $\pi_\lambda=e^{-\beta (H_0-\lambda A)}/Z(\lambda)$ and Hamiltonian evolution. In this case, the Eq.~\eqref{eq:contfdr} reads (as before, we denote by $\phi_{B}(t)$ the response function of observable $B$ under the perturbation $-\lambda(t)A$):
\begin{eqnarray}
\phi_{B}(t) &=& - \frac{\langle \dot B(t)~\Lambda_0 + \Lambda_0 \dot B(t)\rangle_0}{2}\nonumber\\
&=&-\frac{i}{2\hbar}\langle [H_0, B(t)]~\Lambda_0 + \Lambda_0~[H_0, B(t)]\rangle_0\nonumber \\
&=& -\frac{i}{2\hbar}{\rm Tr}\big[ (\Lambda_0~\pi_0+\pi_0~\Lambda_0)~[H_0, B(t)]\big] \nonumber \\
&=& -\frac{i}{2\hbar}{\rm Tr}\big[ [\Lambda_0~\pi_0+\pi_0~\Lambda_0,H_0]~B(t)\big]. \label{eq:phi1}\\ \nonumber
\end{eqnarray}
To proceed further, we need an additional formula for the SLD. For any real function $f(x)$:
\begin{equation}
[H_0-\lambda A,f(H_0-\lambda A)]=0.
\end{equation}
Differentiating this equation with respect to $\lambda$ and setting $\lambda=0$, yields:
\begin{equation}
-[A,f(H_0)]+\left[ H_0, \left.\frac{\partial f(H_0-\lambda A)}{\partial \lambda}\right|_{\lambda=0} \right]=0.
\end{equation}
Particularizing to $f(x)=e^{-\beta x}$ and using the definition of the SLD given by Eq.~\eqref{eq:sld1}  the above equation reduces to:
\begin{equation}
[A,\pi_0]=\frac{1}{2}[H_0,\Lambda_0~\pi_0+\pi_0~\Lambda_0].
\end{equation}
Introducing this last result into Eq.~\eqref{eq:phi1} we finally reach at: 
\begin{equation}
\phi_{B}(t)=\frac{i}{\hbar}{\rm Tr}\big[[A,\pi_0]~B(t)\big]=\frac{i}{\hbar}\langle [B(t),A]\rangle.
\end{equation}
which is the standard Kubo formula. It is interesting to recall that the Kubo formula allows one to express the absorptive part of the susceptibility in terms of a two time correlation {\it in the frequency domain}. This is the so-called quantum FDT, which reads \cite{Cloizeaux}:
\begin{equation}\label{chiC}
\chi''_B(\omega)= \frac{1}{\hbar}\tanh\left(\frac{\beta\hbar\omega}{2}\right)\tilde C_{BA}(\omega),
\end{equation}
where $\tilde C_{BA}(\omega)$ is the Fourier transform of the symmetric correlation function $C_{BA}(t)=\langle B(t)A+AB(t)\rangle_0/2 -\langle B\rangle_0\langle A\rangle_0$. The above equation does not have a simple correlate in the time domain for quantum systems. Only in the classical case, $\hbar\to 0$, Eq.~\eqref{chiC} reduces to:
\begin{equation}\label{chiC2}
\chi''_B(\omega) =\frac{\beta\omega}{2}\,\tilde C_{BA}(\omega),
\end{equation}
which is equivalent to:
\begin{equation}
\phi_B(t) = -{\beta}\,\frac{d}{dt} C_{BA}(t).
\end{equation}
Here we see the clear advantage of the introduction of the SLD conjugated variable $\Lambda_0$: it allows us to express the response function in the time domain as a correlation, {\it both} in the quantum and in the classical case.

\section{The response function of coupled harmonic oscillators}
\label{app C}
Here with the help of the stationary state of the coupled harmonic oscillators, we find the response function of any observable $B$ to the perturbation. 
The covariance matrix (CM) corresponding to the density matrix of Eq.~\eqref{initial-double-rho}, i.e., for a vanishing interaction, is described by the following diagonal matrix:
\begin{align}
\sigma^{\infty}_{_{J=0}}=
\left(
\begin{array}{cccc}
N_1+\frac{1}{2}&0&0&0\\
0&N_2+\frac{1}{2}&0&0\\
0&0&N_1+\frac{1}{2}&0\\
0&0&0&N_2+\frac{1}{2}\\
\end{array}
\right).
\label{eq-stationary-double-uncoupled-cov}
\end{align}
For a non-zero coupling, the CM is not diagonal anymore, as the coupling establishes correlations amongst the two harmonic oscillators. The corresponding CM is given by the following matrix \cite{Campbell2016},
\begin{align}
\sigma^{\infty}=\zeta
\left(
\begin{array}{cccc}
D+N_1+\frac{1}{2}&-\delta C&0&\gamma C\\
-\delta C & D+N_2+\frac{1}{2}&-\gamma C & 0\\
0&-\gamma C & D+N_1+\frac{1}{2}&-\delta C\\
\gamma C & 0 & -\delta C& D+N_2+\frac{1}{2}\\
\end{array}
\right),
\label{eq-two-cloupled-ho-two-baths-covariance}
\end{align}
with $\zeta=\frac{\gamma^2+\delta^2}{4J^2+\gamma^2+\delta^2}$, $D=\frac{2J^2(N_1+N_2+1)}{\gamma^2+\delta^2}$, and $C=\frac{J(N_1-N_2)}{\gamma^2+\delta^2}$. 

Next, we need to identify (i) the SLD associated to $J$, and (ii) the time evolution of the desired observable $B(t)$, under the unperturbed map.
One can find (i) with the help of Eq.~\eqref{eq-two-cloupled-ho-two-baths-covariance}. We know that for such a Gaussian state, $\mathrm{\Lambda}_0$ can be expressed as a linear combination of all the second order moments of the quadratures \cite{monras2013phase,PhysRevA.89.032128}, namely 
\begin{align}
\mathrm{\Lambda}_0&=d_1\left[x_1^2-\left(\sigma^{\infty}_{11}\right)_{J=0}\right]+d_2\left[p_1^2-\left(\sigma^{\infty}_{33}\right)_{J=0}\right]+d_3\left[x_1p_1+p_1x_1\right]\nonumber\\
&+d_4\left[x_2^2-\left(\sigma^{\infty}_{22}\right)_{J=0}\right]+d_5\left[p_2^2-\left(\sigma^{\infty}_{44}\right)_{J=0}\right]+d_6\left[x_2p_2+p_2x_2\right]\nonumber\\
&+c_1~x_1x_2+c_2~x_1p_2+c_3~p_1x_2+c_4~p_1p_2,
\label{eq-SLD-Double-All-Coefficients}
\end{align}
with $d_j$s and $c_j$s being coefficients which are to be determined.
To this end, we make benefit of Eq.~\eqref{eq:staticfdr} of the main text, which states that for any observable $B$ we have:
\begin{align}
\partial_J\left.\right|_{J=0}\average{B}_J=\frac{1}{2}\average{\mathrm{\Lambda}_0B+B\mathrm{\Lambda}_0}_{0}.
\label{eq-lr-Tunneling}
\end{align}
Next, we imply this relation to all of the quadratic observables appearing in \eqref{eq-SLD-Double-All-Coefficients}, i.e., we choose $B\in\{x_1^2,x_2^2,x_1p_1+p_1x_1,\dots\}$. With such choices of $B$, the left hand side of Eq.~\eqref{eq-lr-Tunneling} can be evaluated by taking derivative from the covariance matrix $\sigma^{\infty}$. Moreover, by using the Wick's theorem for Gaussian distributions \cite{Louisell:103059}, one can easily simplify the right hand side and write it down in terms of the elements of $\sigma^{\infty}$ as well. 
We shall start by focusing on the local terms. As an example, for $B=x_1^2$ we have:
\begin{gather}
\frac{1}{2}\average{\mathrm{\Lambda}_0x_1^2+x_1^2\mathrm{\Lambda}_0}_{0}=\left(\partial_J\sigma^{\infty}_{11}\right)_{_{J=0}}~,\nonumber\\
\Rightarrow 2d_1(\sigma^{\infty}_{11})^2_{_{J=0}}-\frac{d_2}{2}=0,
\label{eq-sld-diagonal-x1}
\end{gather}
where we used the fact that $\sigma^{\infty}_{J=0}$, has no off-diagonal terms, as stated by Eq.~\eqref{eq-stationary-double-uncoupled-cov}.
For $B=p_1^2$ one finds a similar equation, with the change of coefficients $d_1\leftrightarrow d_2$.
This implies that, $d_1=d_2=0$. In the same manner, one can find that $d_4=d_5=0$.
For the other two local terms i.e., $B\in\{\frac{1}{2}(x_1p_1+p_1x_1),\frac{1}{2}(x_2p_2+p_2x_2)\}$, Eq.~\eqref{eq-lr-Tunneling} leads to:
\begin{gather}
d_3\left[2(\sigma^{\infty}_{11})_{_{J=0}}(\sigma^{\infty}_{33})_{_{J=0}}+\frac{1}{2}\right]=0,\nonumber\\
d_6\left[2(\sigma^{\infty}_{22})_{_{J=0}}(\sigma^{\infty}_{44})_{_{J=0}}+\frac{1}{2}\right]=0,
\end{gather}
whence, $d_3=d_6=0$ as well. This confirms that the coefficients associated to local observables are zero, i.e., $d_i=0~\forall i$. 
However, the non-local coefficients are non-zero. For $B\in\{x_1x_2,x_1p_2,p_1x_2,p_1p_2\}$, using Eq.~\eqref{eq-lr-Tunneling} yields:
\begin{align}
c_1(\sigma^{\infty}_{11})_{_{J=0}}(\sigma^{\infty}_{22})_{_{J=0}}=(\partial_J\sigma^{\infty}_{12})_{_{J=0}}, ~~~~~~~
c_2(\sigma^{\infty}_{11})_{_{J=0}}(\sigma^{\infty}_{44})_{_{J=0}}=(\partial_J\sigma^{\infty}_{14})_{_{J=0}},\nonumber\\
c_3(\sigma^{\infty}_{33})_{_{J=0}}(\sigma^{\infty}_{22})_{_{J=0}}=(\partial_J\sigma^{\infty}_{32})_{_{J=0}}, ~~~~~~~
c_4(\sigma^{\infty}_{33})_{_{J=0}}(\sigma^{\infty}_{44})_{_{J=0}}=(\partial_J\sigma^{\infty}_{34})_{_{J=0}}.
\end{align}
By using the symmetry in the covariance matrix one can simplify to arrive at:
\begin{align}
c_1=c_4=\left(\frac{\partial_J\sigma^{\infty}_{12}}{\sigma^{\infty}_{11}\sigma^{\infty}_{22}}\right)_{_{J=0}},~~~~~~
c_2=-c_3=\left(\frac{\partial_J\sigma^{\infty}_{14}}{\sigma^{\infty}_{11}\sigma^{\infty}_{44}}\right)_{_{J=0}}.
\end{align}
Replacing in Eq.~\eqref{eq-SLD-Double-All-Coefficients} for the SLD yields:
\begin{align}\label{eq-static-SLD-tunneling}
\mathrm{\Lambda}_0&=c_1(x_1x_2+p_1p_2)+c_2(x_1p_2-p_1x_2)\nonumber\\
&=(c_1+ic_2)a_1^{\dagger}a_2+h.c.
\end{align}
The other key element (ii) is also easy to evaluate for any choice of $B\in\{x_1^2,x_2^2,x_1p_1+p_1x_1,\dots\}$, because the time evolution shall be evaluated under the unperturbed map. 
To begin with, we remind that the time evolution of the ladder operators are easy to find (see \cite{breuer2002theory}, for instance), and read as follow:
\begin{gather}\label{eq-Heisenberg-HO-ladder}
a_{j}(t)=e^{(-i\omega_j-\frac{\gamma_0}{2})t}a_j,~~~~~
a_{j}^{\dagger}(t)=e^{(i\omega_j-\frac{\gamma_0}{2})t}a_j^{\dagger},\nonumber\\
a^{\dagger}_ja_{j}(t)=e^{-\gamma_0 t}a^{\dagger}_ja_j+N(1-e^{-\gamma_0 t}),~~~~~
a_ja^{\dagger}_{j}(t)=e^{-\gamma_0 t}a_ja^{\dagger}_j+(N+1)(1-e^{-\gamma_0 t}).
\end{gather}
Therefore, by using the definition of $x_j$ and $p_j$ quadratures, we find the first moments to evolve as:
\begin{align}\label{eq-time-evolution-global}
x_j(t)&=\frac{1}{\sqrt{2}}(a_j^{\dagger}(t)+a_j(t))=\mathrm{e}^{-\gamma/2 t}\left(\cos(\omega_jt)x_j+\sin(\omega_jt)p_j\right),\nonumber\\
p_j(t)&=\frac{i}{\sqrt{2}}(a_j^{\dagger}(t)-a_j(t))=\mathrm{e}^{-\gamma/2 t}\left(-\sin(\omega_jt)x_j+\cos(\omega_jt)p_j\right).
\end{align}
In addition, the local second moments are:
\begin{align}\label{eq-time-evolution-local}
x_j^2(t)&=e^{-\gamma t}\left(\cos^2(\omega_j t)x_j^2+\sin^2(\omega_j t)p_j^2+\frac{\sin(2\omega_j t)}{2}(x_jp_j+p_jx_j)\right)+(N_j+\tfrac{1}{2})(1-e^{-\gamma t}),\nonumber\\
p_j^2(t)&=e^{-\gamma t}\left(\sin^2(\omega_j t)x_j^2+\cos^2(\omega_j t)p_j^2-\frac{\sin(2\omega_j t)}{2}(x_jp_j+p_jx_j)\right)+(N_j+\tfrac{1}{2})(1-e^{-\gamma t}),\nonumber\\
\left(p_jx_j(t)+x_jp_j(t)\right)&=e^{-\gamma t}\left(\cos(2\omega_j t)\left(x_jp_j+p_jx_j\right)-\sin(2\omega_j t)\left(x_j^2-p_j^2\right)\right).
\end{align}
Note that the evolution keeps locality of the quadratures. On this account, their correlations with the non-local $\Lambda_0$ vanishes at any time. In other words for $B\in\{x_j^2,p_j^2,x_jp_j+p_jx_j\}$ the response function $\phi_B(t)=0$, hence they are blind to the perturbation.\\
On the contrary, for the non-local second moments we have: 
\begin{align}
x_1x_2(t)&=\mathrm{e}^{-\gamma t}\Big(\cos(\omega_1 t)\cos(\omega_2 t)x_1x_2+\cos(\omega_1 t)\sin(\omega_2 t)x_1p_2+\sin(\omega_1 t)\cos(\omega_2 t)p_1x_2+\sin(\omega_1 t)\sin(\omega_2 t)p_1p_2\Big),\nonumber\\
x_1p_2(t)&=\mathrm{e}^{-\gamma t}\Big(-\cos(\omega_1 t)\sin(\omega_2 t)x_1x_2+\cos(\omega_1 t)\cos(\omega_2 t)x_1p_2-\sin(\omega_1 t)\sin(\omega_2 t)p_1x_2+\sin(\omega_1 t)\cos(\omega_2 t)p_1p_2\Big),\nonumber\\
p_1x_2(t)&=\mathrm{e}^{-\gamma t}\Big(-\sin(\omega_1 t)\cos(\omega_2 t)x_1x_2-\sin(\omega_1 t)\sin(\omega_2 t)x_1p_2+\cos(\omega_1 t)\cos(\omega_2 t)p_1x_2+\cos(\omega_1 t)\sin(\omega_2 t)p_1p_2\Big),\nonumber\\
p_1p_2(t)&=\mathrm{e}^{-\gamma t}\Big(\sin(\omega_1 t)\sin(\omega_2 t)x_1x_2-\sin(\omega_1 t)\cos(\omega_2 t)x_1p_2-\cos(\omega_1 t)\sin(\omega_2 t)p_1x_2+\cos(\omega_1 t)\cos(\omega_2 t)p_1p_2\Big),
\label{second-moments-time}
\end{align}
which are all non-local, hence we expect a non-zero response for them. To examine this, let us focus on $B=x_1x_2$. By replacing Eqs.~\eqref{second-moments-time} and \eqref{eq-static-SLD-tunneling} into the definition of the response function, Eq.~\eqref{eq:contfdr}, we obtain:
\begin{align}
\phi_{x_1x_2}(t)&={\rm Corr}\big(\Lambda_0,x_1x_2(t)\big)_0\nonumber\\
&=\mathrm{e}^{-\gamma t}\big( c_1(\sigma^{\infty}_{11})_{_{J=0}}(\sigma^{\infty}_{22})_{_{J=0}}\left[\cos(\omega_1 t)\cos(\omega_2 t)+\sin(\omega_1 t)\sin(\omega_2 t)\right]\nonumber\\
&~~~~~~~~~+c_2(\sigma^{\infty}_{11})_{_{J=0}}(\sigma^{\infty}_{22})_{_{J=0}}\left[\cos(\omega_1 t)\sin(\omega_2 t)-\sin(\omega_1 t)\cos(\omega_2 t)\right]\big)\nonumber\\
&=\frac{N_2-N_1}{\gamma^2+\delta^2}\mathrm{e}^{-\gamma t}\big(\delta\cos\delta t + \gamma\sin\delta t\big)\nonumber\\
&=\frac{N_2-N_1}{\sqrt{\gamma^2+\delta^2}}\mathrm{e}^{-\gamma t}\cos(\delta t-\theta),
\end{align}
where we define $\theta=\arctan\gamma/\delta$. By using the same strategy, it is easy to verify that the response function of the other non-local observables read as:
\begin{align}
\phi_{x_1p_2}(t)&=\frac{N_1-N_2}{\gamma^2+\delta^2}\mathrm{e}^{-\gamma t}\big(\delta\sin\delta t - \gamma\cos\delta t\big)
=\frac{N_1-N_2}{\sqrt{\gamma^2+\delta^2}}\mathrm{e}^{-\gamma t}\sin(\delta t-\theta),\nonumber\\
\phi_{p_1x_2}(t)&=\frac{N_2-N_1}{\gamma^2+\delta^2}\mathrm{e}^{-\gamma t}\big(\delta\sin\delta t - \gamma\cos\delta t\big)
=\frac{N_2-N_1}{\sqrt{\gamma^2+\delta^2}}\mathrm{e}^{-\gamma t}\sin(\delta t-\theta),\nonumber\\
\phi_{p_1p_2}(t)&=\frac{N_2-N_1}{\gamma^2+\delta^2}\mathrm{e}^{-\gamma t}\big(\delta\cos\delta t + \gamma\sin\delta t\big)
=\frac{N_2-N_1}{\sqrt{\gamma^2+\delta^2}}\mathrm{e}^{-\gamma t}\cos(\delta t-\theta).
\end{align}
\twocolumngrid

\bibliographystyle{apsrev4-1}
\bibliography{Refs}

\end{document}